\documentclass{article}

\usepackage{amsmath,amssymb,mathrsfs}
\usepackage{graphicx}

\usepackage{url} 
\usepackage{lineno}

\newcommand{\xpos}{\boldsymbol{x}}
\newcommand{\ypos}{\boldsymbol{y}}

\begin{document}

\title{Tunable high-resolution synthetic aperture radar
  imaging}

\author{Arnold D.~Kim\footnotemark[1]  \ and Chrysoula Tsogka\footnotemark[1]}
\renewcommand{\thefootnote}{\fnsymbol{footnote}}

\footnotetext[1]{Department of Applied Mathematics, University of
  California, Merced, 5200 North Lake Road, Merced, CA 95343, USA}
\maketitle

%
%
%
%

\begin{abstract}
  We have recently introduced a modification of the multiple signal
  classification (MUSIC) method for synthetic aperture radar. This
  method depends on a tunable, user-defined parameter,
  $\epsilon$, that allows for quantitative high-resolution imaging. It
  requires however, relative large single-to-noise ratios (SNR) to
  work effectively. Here, we first identify the fundamental mechanism
  in that method that produces high-resolution images. Then we
  introduce a modification to Kirchhoff Migration (KM) that uses the same
  mechanism to produces tunable, high-resolution images. This modified
  KM method can be applied to low SNR measurements. We show simulation
  results that demonstrate the features of this method.
\end{abstract}

\section{Introduction}

In synthetic aperture radar (SAR) a single transmitter/receiver on a
platform is used to probe an imaging region of interest by emitting
pulses and then record the subsequent echoes as the platform moves
along a flight path.
SAR imaging methods use these measurements to reconstruct images of
targets in the imaging region of interest.

The traditional SAR image is formed by evaluating the data at each
measurement location at the travel time that it takes for the waves to
propagate from the platform location to a point in the imaging region
on the ground and back. This imaging method is called Kirchhoff
Migration (KM). The resolution of the image produced by KM has a range
resolution that is $O(c/B)$ with $c$ denoting the wave speed, and $B$
denoting the system bandwidth, and a cross-range resolution that is
$O(\lambda_{0} L/a)$ with $\lambda_{0}$ denoting the central
wavelength, $L$ denoting the characteristic distance from the platform
to the imaging region, and $a$ denoting the length of the synthetic
aperture~\cite{cheney2009fundamentals}.

The authors have recently developed a modification to the multiple
signal classification method (MUSIC) for SAR imaging~\cite{MUSIC4SAR}. 
This method includes a user-defined parameter, which we call $\epsilon$, that
allows for tunable, high-resolution images. Using this method, we
obtain a range resolution that is $O(\sqrt{\epsilon} (c/B) (L/R) )$
with $R$ denoting the range distance from the flight path to the
center of the imaging region, and a cross-range resolution of
$O(\sqrt{\epsilon} (c/B) (L/a) )$. The key here is that the
user-defined parameter $\epsilon$ can be made arbitrarily small which,
in turn, allows for tunable high-resolution quantitative
images. However, this method is limited to measurements with
relatively high signal-to-noise ratios (SNRs). Otherwise, one cannot
reliably separate signal from noise, which is a key step in this
method. The quantitative recovery of reflectivity information suffers
most when the SNR becomes too low, but image resolution also suffers.

In many practical applications, SAR data are very
noisy~\cite{doerry}. For these problems, the modification to MUSIC
will not be effective because the SNR is not sufficiently high. For
this reason, we seek to develop a new method that retains the tunable
high-resolution feature of this method, but can accommodate
measurements with low SNR. To do this, we first re-evaluate the
modification to MUSIC and identify the fundamental mechanism leading
to tunable high-resolution. This mechanism involves a simple rational
transformation. We form the same rational transformation of the
normalized KM image and obtain a tunable high-resolution version of
the KM method. This method requires no additional computations beyond
those required for KM. The result of this modification is a method
with a range resolution that is $O(\sqrt{\epsilon} c/B)$ and a
cross-range resolution that is $O(\sqrt{\epsilon} \lambda_{0} L/a)$.
  
The remainder of this paper is as follows. In Section \ref{sec:sar} we
give a brief description of synthetic aperture radar imaging, our
assumptions, and define the measurements. In Section \ref{sec:music}
we briefly review the recent modification to MUSIC and identify the
fundamental mechanism allowing for tunable high-resolution
imaging. Using the insight gained from identifying this fundamental
mechanism, we introduce a modification to KM in Section
\ref{sec:modifiedKM}. We show several simulation results in Section
\ref{sec:results} that demonstrate the features of the modified KM
method. Section \ref{sec:conclusions} contains our conclusions.

\section{Synthetic aperture radar imaging}
\label{sec:sar}

We consider synthetic aperture radar (SAR) imaging in which a single
transmitter/receiver on a moving platform emits and records signals
\cite{cheney2001mathematical, cheney2009fundamentals,
  moreira2013tutorial}. A full set of measurements corresponds to a
suite of experiments at several locations along the flight path. Let
$f(t)$ denote the broadband pulse emitted and let $d(s,t)$ denote the
data recorded which depends on the slow time $s$ parameterizing the
flight path of the platform, $\boldsymbol{r}(s)$, and the fast time
$t$ corresponding to the round-trip travel time between the platform
and the imaging scene. Here, we assume the start-stop approximation in
which the change in displacement between the targets and the platform
is negligibly small compared to the travel time it takes for the pulse
emitted to propagate to the imaging scene and return as echoes.

\begin{figure}[t]
  \centering
  \includegraphics[width=0.65\textwidth]{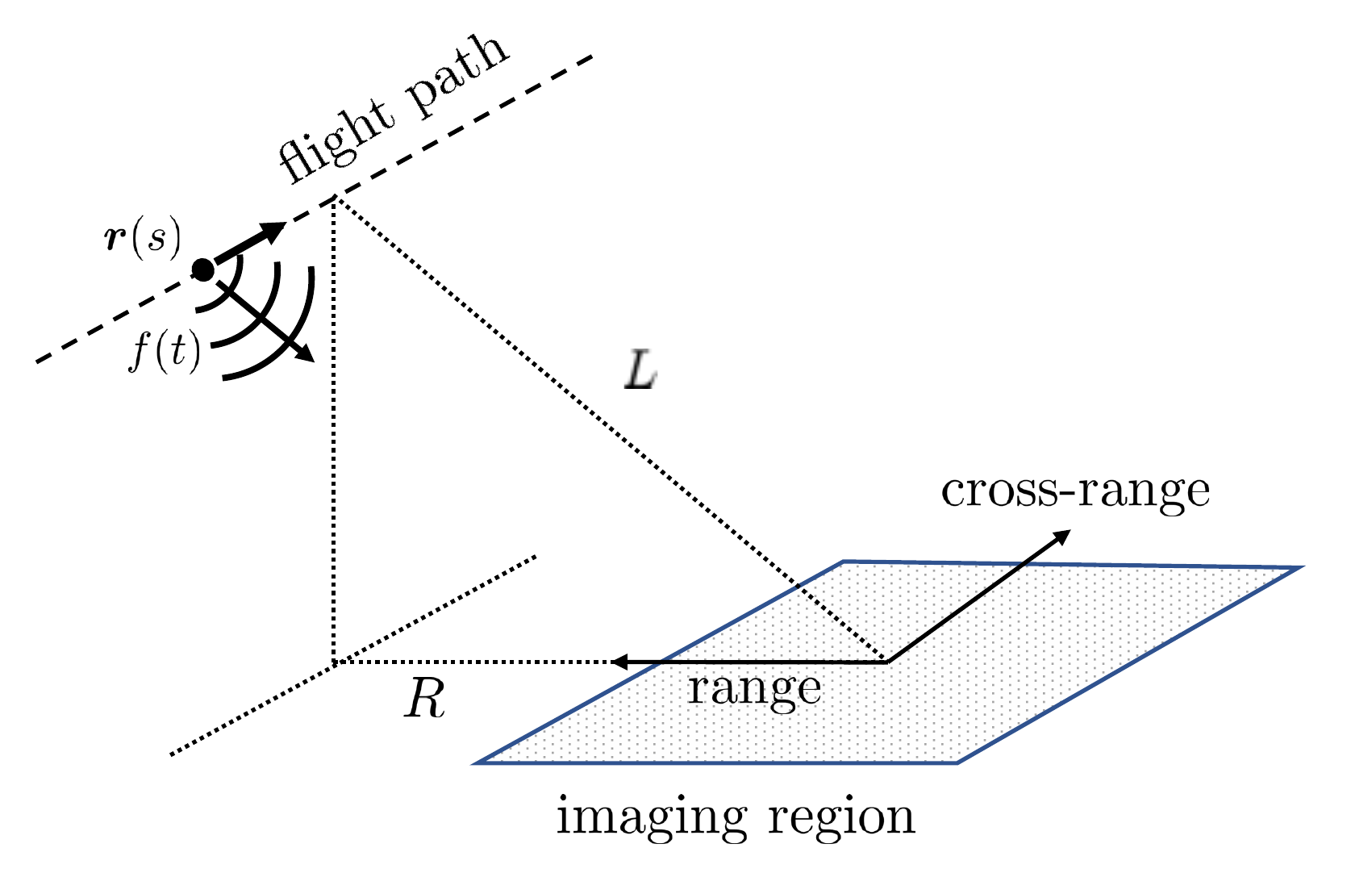}
  \caption{Setup for synthetic aperture radar imaging.}
  \label{fig:schematic}
\end{figure}

Figure~\ref{fig:schematic} shows an illustration of SAR imaging. Here,
we see that range is the coordinate obtained by projecting the vector
that connects the center of the imaging region to the central platform
location onto the imaging plane, and cross-range is the coordinate
orthogonal to range. Denoting the size of the synthetic aperture by
$a$ and the system bandwidth by $B$, the typical resolution of the
imaging system is $O((c/B) (L/R))$ in range and $O(\lambda L/a)$ in
cross-range. Here $c$ is the speed of light and $\lambda$ the
wavelength corresponding to the central frequency while $L$ denotes
the distance between the platform and the imaging region and $R$ the
offset in range.
 
With the start-stop approximation, measurements are taken at $N$
discrete values of $s$, corresponding to $d(s_{n},t)$ for
$n = 1, \cdots, N$.  Next, 
we compute the discrete Fourier transform of digitally sampled values
of $d(s_{n},t)$ in $t$ denoted by $d_{n}(\omega_{m})$ for
$m = 1, \cdots, 2M-1$. The full set of measurements is the
$2M-1 \times N$ matrix $D$ whose columns are

\begin{equation}
  \mathbf{d}_{n} = 
  \begin{bmatrix}
    d_{n}(\omega_{1}) \\ d_{n}(\omega_{2}) \\ \vdots \\
    d_{n}(\omega_{2M-1})
  \end{bmatrix}, \quad n = 1, \cdots, N.
  \label{eq:Dmatrix}
\end{equation}
The objective in SAR imaging is to identify and locate targets in an
imaging region using these data.

\section{MUSIC for SAR}
\label{sec:music}

The authors have recently extended MUSIC for SAR resulting in a
tunable, quantitative high-resolution imaging method~\cite{MUSIC4SAR}.
The main limitation of this method is that it requires sufficiently
high SNR to distinguish between signal and noise subspaces. We review
this method here and identify the key mechanism that leads to tunable
high-resolution images.

The key to applying MUSIC to SAR is a reorganization of the data
matrix $D$ which we explain below. Consider the $n$th column of
$D$ given in \eqref{eq:Dmatrix}. Applying the Prony
method~\cite{prony1795essai} to this vector of length $2M-1$ yields
the following $M \times M$ matrix,

\begin{equation}
  D_{n} = \begin{bmatrix} 
    d_{n}(\omega_{1}) & d_{n}(\omega_{2}) & \cdots &
    d_{n}(\omega_{M}) \\
    d_{n}(\omega_{2}) & d_{n}(\omega_{3}) & \cdots &
    d_{n}(\omega_{M+1}) \\
    \vdots & \vdots & \ddots & \vdots \\
    d_{n}(\omega_{M}) & d_{n}(\omega_{M+1}) & \cdots &
    d_{n}(\omega_{2M-1})
  \end{bmatrix}.
  \label{eq:Pronyfied}
\end{equation}
In this rearrangement, the first column is the truncation of
$\mathbf{d}_{n}$ to its first $M$ entries. Subsequent columns are
sequential upward shifts of $\mathbf{d}_{n}$ truncated to its first
$M$ entries.

When there is no measurement noise and the frequencies are sampled at
a fixed rate, \eqref{eq:Pronyfied} can be factorized as a sum of outer
products, each corresponding to an individual point
target~\cite{MUSIC4SAR}. Suppose there are $K$ point targets in the
imaging region located at positions $\ypos_{k}$ with reflectivities
$\rho_{k}$ for $k = 1, \cdots, K$. It follows that

\begin{equation}
  D_{n} = \sum_{k = 1}^{K} s_{k}^{(n)} \mathbf{u}_{k}^{(n)}
  {\mathbf{v}_{k}^{(n)}}^{H},
\end{equation}
with $s_{k}^{(n)} = \rho_{k}/(4 \pi | \xpos_{n} - \ypos_{k}|)^{2}$,
\begin{equation}
  \mathbf{u}_{k}^{(n)} = 
  \begin{bmatrix}
    e^{\mathrm{i} 2 \omega_{1} | \xpos_{n} - \ypos_{k} |/c} \\
    e^{\mathrm{i} 2 \omega_{2} | \xpos_{n} - \ypos_{k} |/c} \\
    \vdots \\
    e^{\mathrm{i} 2 \omega_{M} | \xpos_{n} - \ypos_{k} |/c}
  \end{bmatrix},
  \quad
  \text{and}
  \quad
  \mathbf{v}_{k}^{(n)} =
    \begin{bmatrix}
      1 \\
      e^{-\mathrm{i} 2 \Delta \omega | \xpos_{n} - \ypos_{k} |/c} \\
      \vdots \\
      e^{-\mathrm{i} 2 (M-1) \Delta \omega | \xpos_{n} -
        \ypos_{k} |/c}
    \end{bmatrix}.
    \label{eq:uv}
\end{equation}
Here, $\xpos_{n}$ denotes the platform location at the $n$th
measurement, $\omega_{m} = \omega_{1} + (m-1)\Delta\omega$ for
$m = 1, \cdots, M$, and $c$ denotes the wave speed.

Assuming that $K < M$, we find that the column space of $D_{n}$ is
given by

\begin{equation}
  \mathscr{S} = \text{span}\{ \mathbf{u}_{1}^{(n)}, \cdots,
  \mathbf{u}_{K}^{(n)} \},
\end{equation}
which corresponds to the signal subspace. Upon computing the singular
value decomposition $D_{n} = U \Sigma V^{H}$, we find that the first
$K$ columns of $U$ give an orthonormal basis for $\mathscr{S}$. Let
$\tilde{U}$ be the $M \times K$ matrix corresponding to the first $K$
columns of $U$. The projection onto the signal subspace is then
$P_{\text{signal}} = \tilde{U} \tilde{U}^{H}$. The noise subspace is
the orthogonal complement to the signal subspace. The projection onto
the noise subspace is
$P_\text{noise} = I_{M} - \tilde{U} \tilde{U}^{H}$ with $I_{M}$
denoting the $M \times M$ identity matrix.

Now suppose we wish to test if a search point $\ypos$ somewhere in the
imaging region corresponds to a target. We introduce the illumination
vector,

\begin{equation}
  \mathbf{a}_{n}(\ypos) =   \begin{bmatrix}
    e^{\mathrm{i} 2 \omega_{1} | \xpos_{n} - \ypos |/c} \\
    e^{\mathrm{i} 2 \omega_{2} | \xpos_{n} - \ypos |/c} \\
    \vdots \\
    e^{\mathrm{i} 2 \omega_{M} | \xpos_{n} - \ypos |/c}
  \end{bmatrix}.
  \label{eq:illumination}
\end{equation}
In traditional MUSIC we form an image by testing if this illumination
vector lies in the noise subspace. For this case, we compute
$\| P_{\text{noise}} \mathbf{a}_{n}(\ypos) \|^{2}$ for
$n = 1, \cdots, N$, and form an image by plotting

\begin{equation}
  I(\ypos) = \left[ \sum_{n = 1}^{N} \| P_{\text{noise}}
    \mathbf{a}_{n}(\ypos) \|^{2} \right]^{-1}.
\end{equation}
The basic idea behind this formation of an image is that when $\ypos$
corresponds to a target location, the projection onto the noise
subspace will be zero, so plotting the reciprocal will yield a
singularity there.

In the recent modifications to MUSIC \cite{gonzalez2021quantitative,
  MUSIC4SAR}, both $P_{\text{noise}}$ and $P_{\text{signal}}$ are
used. The motivating idea in those methods is that $P_{\text{noise}}$
is used to identify and locate targets and that $P_{\text{signal}}$ is
used to deliver quantitative information about the target. However,
there is a more generally useful reason to include both
projections. By considering both $P_{\text{noise}}$ and
$P_{\text{signal}}$, one can introduce a continuous weighting in terms
of a user-defined parameter $\epsilon > 0$ according to

\begin{align}
  I_{\epsilon}(\ypos)
  &= \left[ \sum_{n = 1}^{N} \epsilon^{-1} \|
    P_{\text{noise}} \mathbf{a}_{n}(\ypos) \|^{2} + \|
    P_{\text{signal}} \mathbf{a}_{n}(\ypos) \|^{2} \right]^{-1}
    \nonumber\\
  &= \epsilon \left[ \sum_{n = 1}^{N} \left( 1 -
    (1 - \epsilon ) \frac{\left| \tilde{U}^{H}
    \mathbf{a}_{n}(\ypos) \right|^{2}}{\| \mathbf{a}_{n}(\ypos)
    \|^{2}} \right) \| \mathbf{a}_{n}(\ypos) \|^{2} \right]^{-1},
  \label{eq:functional}
\end{align}
where we have resubstituted our expressions for $P_{\text{noise}}$ and
$P_{\text{signal}}$ into \eqref{eq:functional}.

For a single target located at $\ypos_{0}$, we have found in a
neighborhood about $\ypos_{0}$ that

\begin{equation}
  \frac{\left| \tilde{U}^{H} \mathbf{a}_{n}(\ypos) \right|^{2}}{\|
    \mathbf{a}_{n}(\ypos) \|^{2}} \approx 1 - \beta^{2} | \ypos -
  \ypos_{0} |^{2}
\end{equation}
for some $\beta$~\cite{MUSIC4SAR}. Consider the simple function
$f(x) = 1 - \beta^{2} x^{2}$. It assumes its maximum value of $1$ on
$x = 0$. It assumes its full-width/half-maximum (FWHM) on
$\Delta x^{\ast} = \pm 1/(\sqrt{2} \beta)$. Now consider the function,

\begin{equation}
  F_{\epsilon}(x) = \frac{\epsilon}{1 - (1 - \epsilon) f(x)}.
  \label{eq:one-over}
\end{equation}
This function also assumes its maximum value of $1$ on $x = 0$, but it
assumes its FWHM on

\begin{equation}
  \Delta x^{\ast} = \pm \frac{\sqrt{\epsilon}}{\beta \sqrt{1 -
      \epsilon}} = \frac{\sqrt{\epsilon}}{\beta} + O(\epsilon^{3/2}),
  \quad \epsilon \to 0.
\end{equation}
Thus for a function $f(x)$ with some FWHM, we can instead consider the
function $F_{\epsilon}(x)$ given in \eqref{eq:one-over} with the same
FWHM, but scaled by $\sqrt{\epsilon}$.  It is through this simple
mechanism 
that we have
determined that this generalization of MUSIC for SAR has a range resolution that is
$O(\sqrt{\epsilon} (c/B) (L/R) )$ and a cross-range resolution that is
$O(\sqrt{\epsilon} (c/B) (L/a) )$.

The key point is the factor of $\sqrt{\epsilon}$ since $\epsilon$ can
be made arbitrarily small.  This tunable resolution can be used
advantageously when forming images of targets. For example, the value
of $\epsilon$ can be varied by the user based on the resolution of the
mesh used to sample the imaging region. Upon identifying subregions
where an individual target is located, one can then change the value
of $\epsilon$ to obtain a very high resolution image of the target on
a finer mesh.

\section{Modified Kirchhoff Migration}
\label{sec:modifiedKM}

The traditional method for SAR imaging uses Kirchhoff Migration
(KM). This imaging method is robust to measurement noise. In fact, it
is not restricted like MUSIC to problems where the signal subspace can
be reliably distinguished from the noise subspace. For that reason, KM
is robust for problems with very low SNR. Here, we give a modification
of KM based off of what we have done for MUSIC to yield a KM imaging
method with tunable resolution.

Using the data defined in \eqref{eq:Dmatrix}, we form the KM image
through evaluation of

\begin{equation}
  I^{\text{KM}}(\ypos) = \sum_{n = 1}^{N} \sum_{m = 1}^{2M-1}
  d_{n}^{\ast}(\omega_{m}) e^{\mathrm{i} 2 \omega_{m} | \xpos_{n} -
      \ypos |/ c},
  \label{eq:KM}
\end{equation}
with $d_{n}^{\ast}(\omega_{m})$ denoting the complex conjugate of
$d_{n}(\omega_{m})$. It is well known that the resolution of
$I^{\text{KM}}$ is $O(c/B)$ in range and $O(\lambda L/a)$ in
cross-range. 

The inner sum over frequencies in \eqref{eq:KM} is an inner product,
so it can be interpreted as a projection of the illumination vector

\begin{equation}
  \mathbf{a}_{n}(\ypos) = \left[ e^{\mathrm{i} 2 \omega_{1} | \xpos_{n} -
    \ypos |}, \cdots, e^{\mathrm{i} 2 \omega_{2M-1} | \xpos_{n} -
    \ypos |} \right]^{T},
\end{equation}
onto the data vector
\begin{equation}
  \mathbf{d}_{n} = \left[ d_{n}(\omega_{1}), \cdots,
    d_{n}(\omega_{2M-1})  \right]^{T},
\end{equation}
or
\begin{equation}
  I^{\text{KM}}(\ypos) = \sum_{n = 1}^{N} \mathbf{d}_{n}^{H}
  \mathbf{a}_{n}(\ypos).
\end{equation}
Let $\mathbf{d}_{n} = \hat{\mathbf{d}}_{n} \| \mathbf{d}_{n} \|$ and
$\mathbf{a}_{n}(\ypos) = \hat{\mathbf{a}}_{n}(\ypos) \|
\mathbf{a}_{n}(\ypos) \|$. Just as we have done for MUSIC, we form a
tunable KM imaging functional through evaluation of
\begin{equation}
  I^{\text{KM}}_{\epsilon}(\ypos) = \epsilon \left| \sum_{n = 1}^{N}
    \left( 1 - ( 1 - \epsilon ) \hat{\mathbf{d}}_{n}^{H}
          \hat{\mathbf{a}}_{n}(\ypos) \right)  \right|^{-2}.
    \label{eq:modifiedKM}
\end{equation}
This modification of KM will yield an image with a tunable resolution
that scales as $\sqrt{\epsilon}$. We expect that this imaging method
will have a range resolution that is $O(\sqrt{\epsilon} c/B)$ and a
cross-range resolution that is $O(\sqrt{\epsilon} \lambda_{0}
L/a)$. Since $\epsilon$ is a user-defined parameter, it can be made
arbitrarily small to produce a very high-resolution image.

The mechanism through which \eqref{eq:modifiedKM} achieves high
resolution requires that
$\hat{\mathbf{d}}_{n}^{H} \hat{\mathbf{a}}_{n}(\ypos) = 1$ on target
locations. When considering an imaging region with multiple targets,
this requirement may not be met exactly on all target locations.
Therefore, \eqref{eq:modifiedKM} may not clearly identify some of the
targets. However, if \eqref{eq:modifiedKM} is used in a sub-region
containing a single target, it will produce tunable high-resolution
images of that individual target.

\section{Numerical results}
\label{sec:results}

We use numerical simulations to test and evaluate the imaging method
presented above. We used the following values for the system
parameters which are based on the GOTCHA data
set~\cite{casteel2007challenge}. We set the origin of a coordinate
system in the middle of a $4.8\, \text{m} \times 4.8\, \text{m}$
imaging region situated at ground level, $z = 0$. The $x$-coordinate
is cross-range and the $y$-coordinate is range.  The coordinates along
the linear flight path where measurements are taken are
$\xpos_{n} = ( x_{n}, R, H )$ with $x_{n} = -a/2 + a (n-1)/(N-1)$ for
$n = 1, \cdots, N$.  Here, the range of the flight path is
$R = 7.10\, \text{km}$ and the height is $H = 7.30\, \text{km}$, so
that $L = \sqrt{H^{2} + R^{2}} = 10.18\, \text{km}$.  The length of
the synthetic aperture is $a = 0.13\, \text{km}$. The central
frequency is $f_{0} = 9.6\, \text{GHz}$ and the bandwidth is
$B = 622\, \text{MHz}$. Using $c = 3 \times 10^{8}\, \text{m/s}$, we
find that the central wavelength is $\lambda_{0} = 3.12\,
\text{cm}$. We use $M = 31$ equi-spaced frequencies over the bandwidth
and $N = 124$ spatial measurements.

\subsection{Single target}

\begin{figure}[htb]
  \centering
  \includegraphics[width=0.48\linewidth]{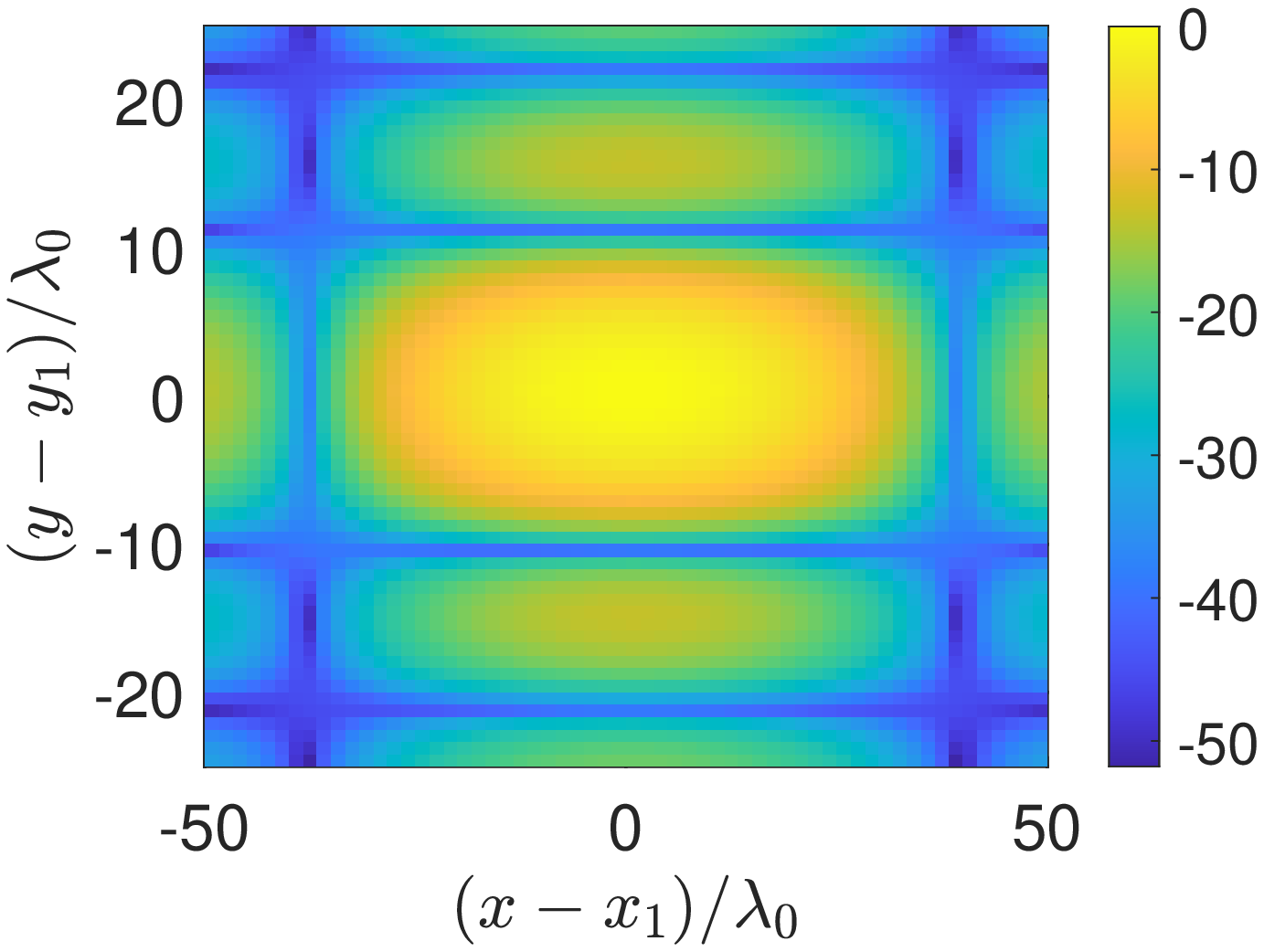}
  \hfill
  \includegraphics[width=0.48\linewidth]{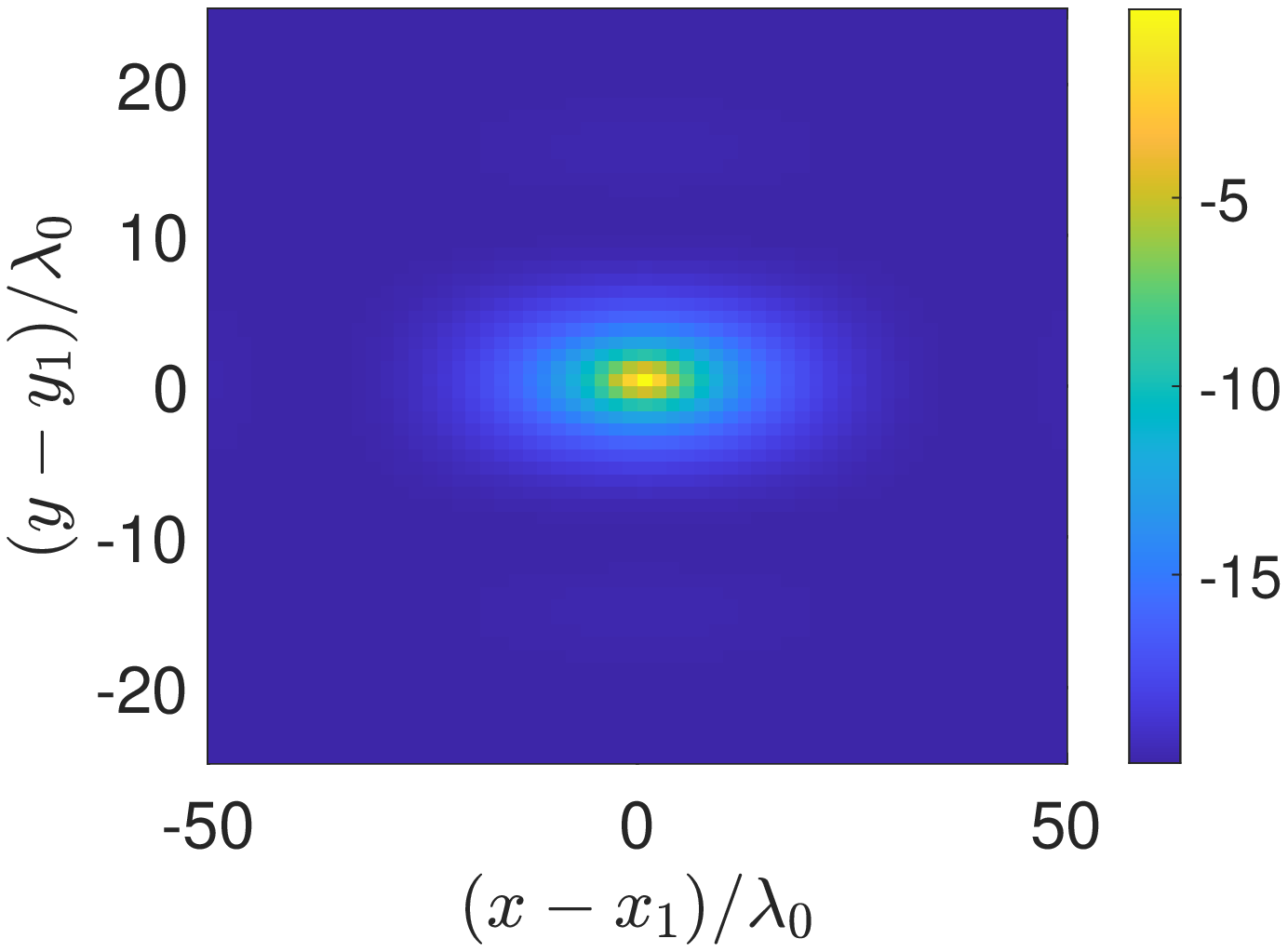}
  \caption{(Left) Image formed by KM centered on the target location,
    normalized by its peak value and plotted in dB
    ($10 \log_{10}$-scale) with $\text{SNR} = 15.3989\,
    \text{dB}$. (Right) Image formed using modified KM with
    $\epsilon = 10^{-4}$ for the same measurement data.}
  \label{fig:1target}
\end{figure}

We first apply the imaging method to a single point target located at
$\ypos_{0} = (1\, \text{m}, 1\, \text{m}, 0)$. We added measurement
noise to the measured signal so that
$\text{SNR} = 15.3989\, \text{dB}$. The image resulting from applying
KM given in \eqref{eq:KM} appears in the left plot and the image
resulting from applying the modified KM given in \eqref{eq:modifiedKM}
appears in the right plot of Fig.~\ref{fig:1target}. For both plots,
the image was normalized by its peak value and plotted in dB
($10 \log_{10}$-scale). Additionally, the images are centered about
the true target location.

Using the parameter values for this problem, we find that
$c/B \approx 15\, \lambda_{0}$ and
$\lambda_{0} L/a \approx 79 \lambda_{0}$ corresponding to the range
($x$) and cross-range ($y$) resolutions, respectively. The left plot
shows a peak at the target location whose full-width/half-maximum
(FWHM) corresponds to those resolutions estimates. Additionally, we
see the familiar side-lobes that appear in KM images.

To form the image appearing the right plot of Fig.~\ref{fig:1target},
we have used $\epsilon = 10^{-4}$ in the modified KM image. This image
has a sharper peak at the target location with no apparent
side-lobes. It has a much higher resolution than the KM image with
less artifacts.

\begin{figure}[htb]
  \centering
  \includegraphics[width=0.32\linewidth]{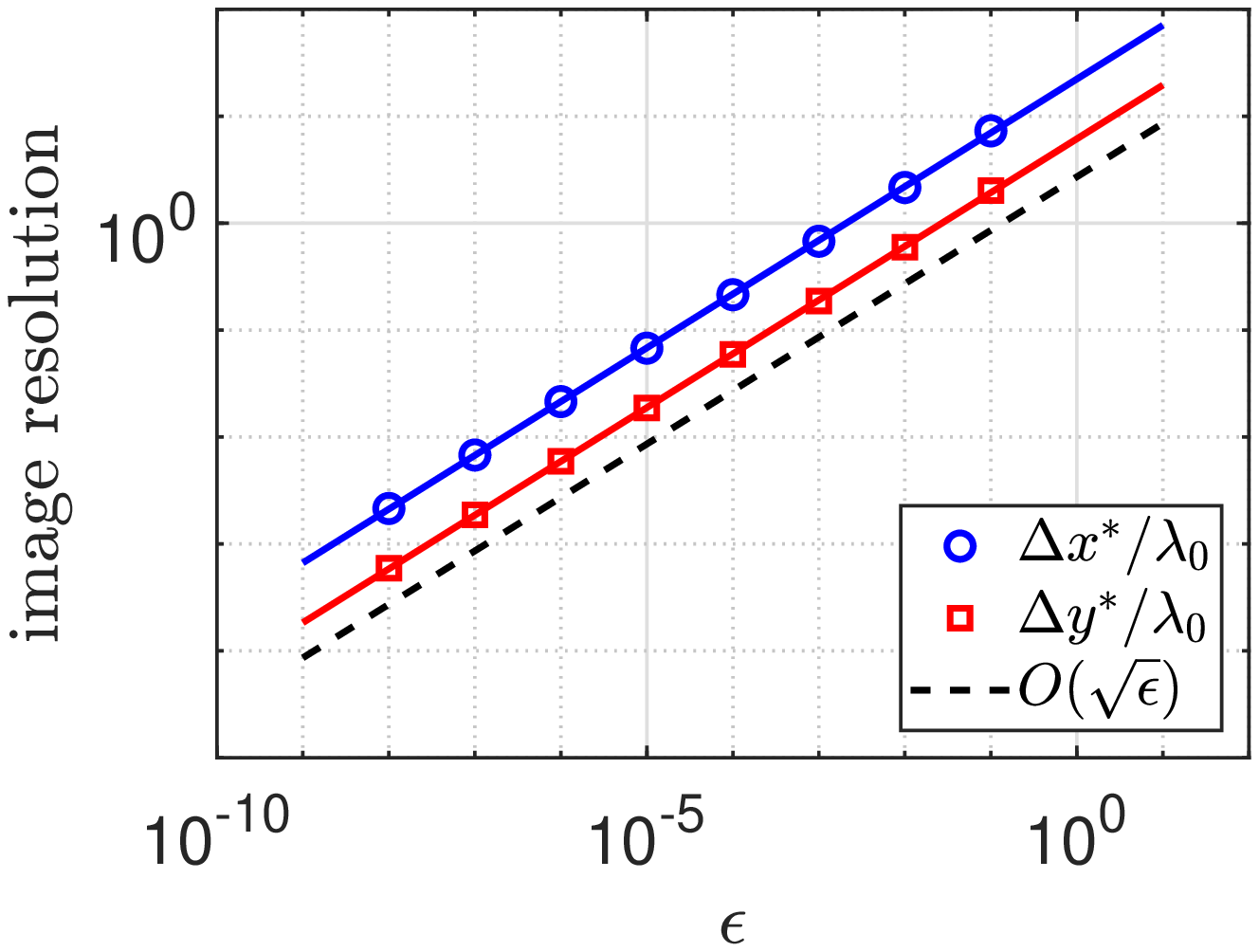}
  \hfill
  \includegraphics[width=0.32\linewidth]{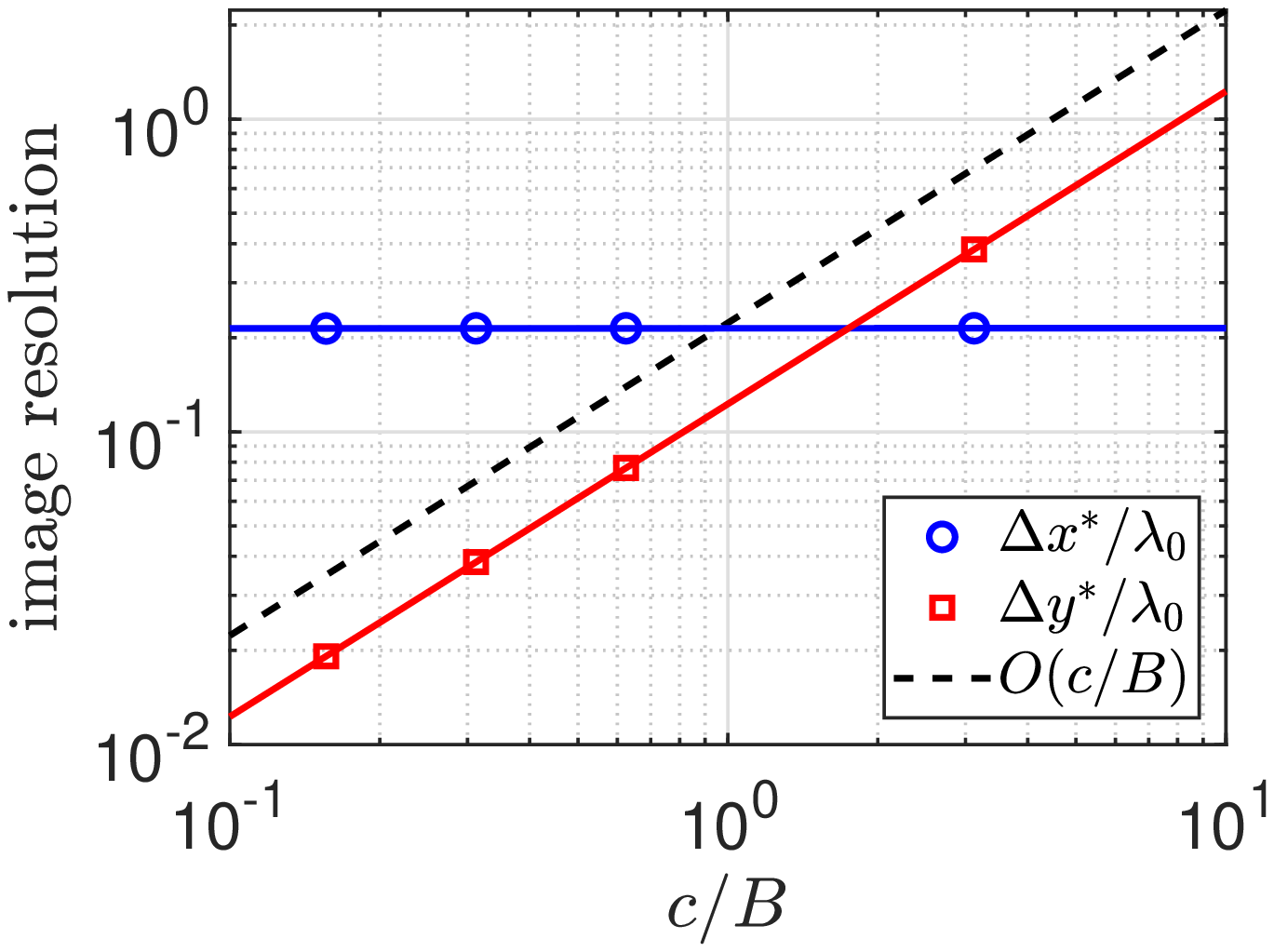}
  \hfill
  \includegraphics[width=0.32\linewidth]{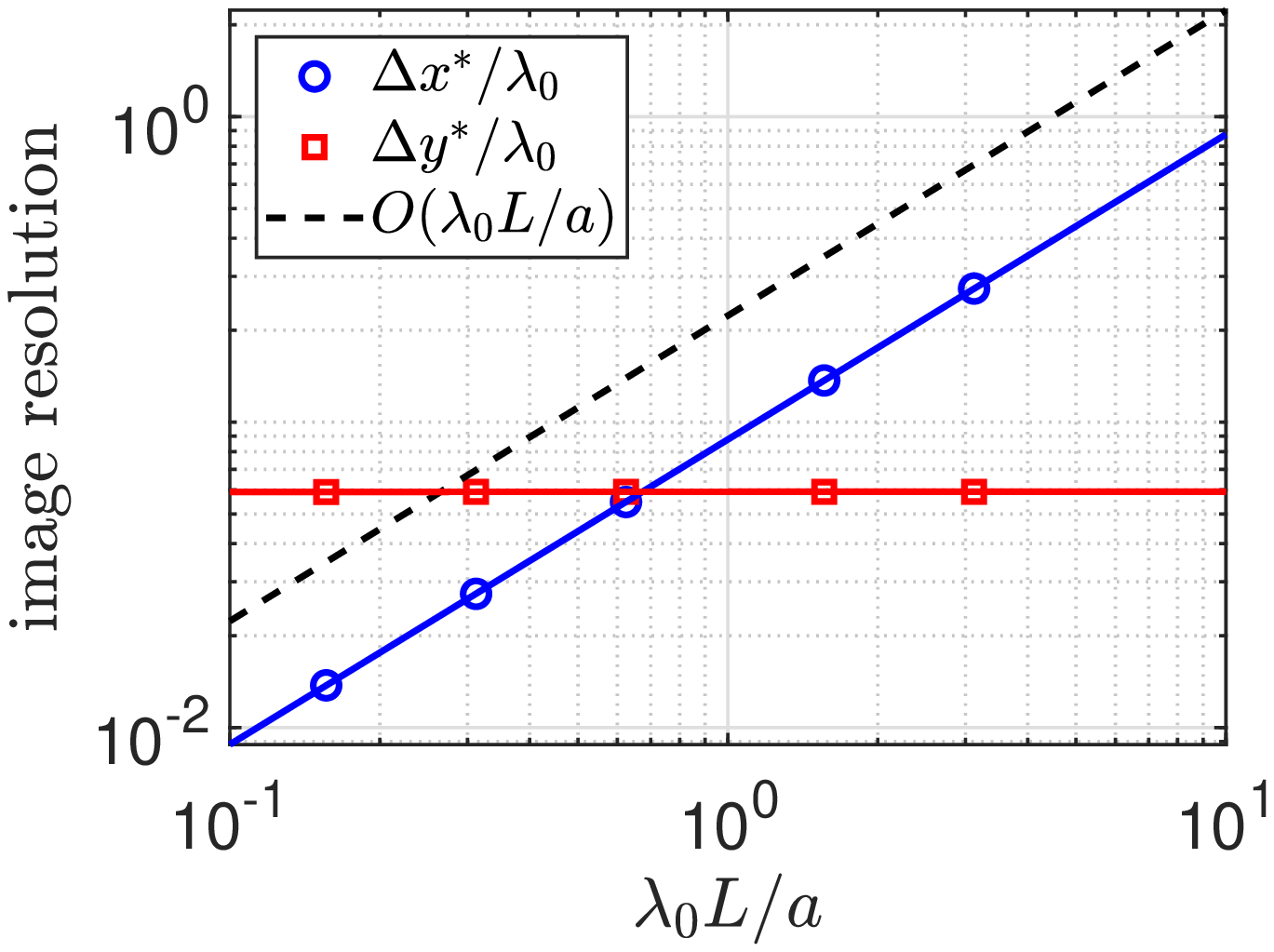}
  \caption{Computed full-width/half-maximum (FWHM) values of images
    produced with modified KM as a function of $\epsilon$ (left),
    $c/B$ (center), and $\lambda_{0} L/a$ (right). }
  \label{fig:1target-resolution}
\end{figure}

We expect that the image resolution of modified KM to be
$O(\sqrt{\epsilon} c/B)$ in range and
$O(\sqrt{\epsilon} \lambda_{0} L/a)$ in cross-range. Using numerical
simulations, we estimate the FWHM in range and cross-range for
noiseless data for the same single target. The image resolution
results are shown in Fig.~\ref{fig:1target-resolution}. The plots in
Fig.~\ref{fig:1target-resolution} show the computed FHWM in range
($\Delta x^{\ast}/\lambda_{0}$) as blue circles, and in cross-range
($\Delta y^{\ast}/\lambda_{0}$) as red squares. The left plot in
Fig.~\ref{fig:1target-resolution} shows the FWHM results as a function
of $\epsilon$ with $c/B$ and $\lambda_{0} L/a$ fixed. The center plot
shows FWHM results as a function of $c/B$ with $\epsilon = 10^{-4}$
and $\lambda_{0} L/a$ fixed. The right plot shows FWHM results as a
function of $\lambda_{0} L/a$ with $\epsilon = 10^{-4}$ and $c/B$
fixed. All three of these plots show linear fits to the data as solid
curves. The black dashed curve shows the expected behavior of the
image resolution.  These results verify the resolution estimates for
modified KM.

\subsection{Multiple targets}

\begin{figure}[htb]
  \centering
  \includegraphics[width=0.48\linewidth]{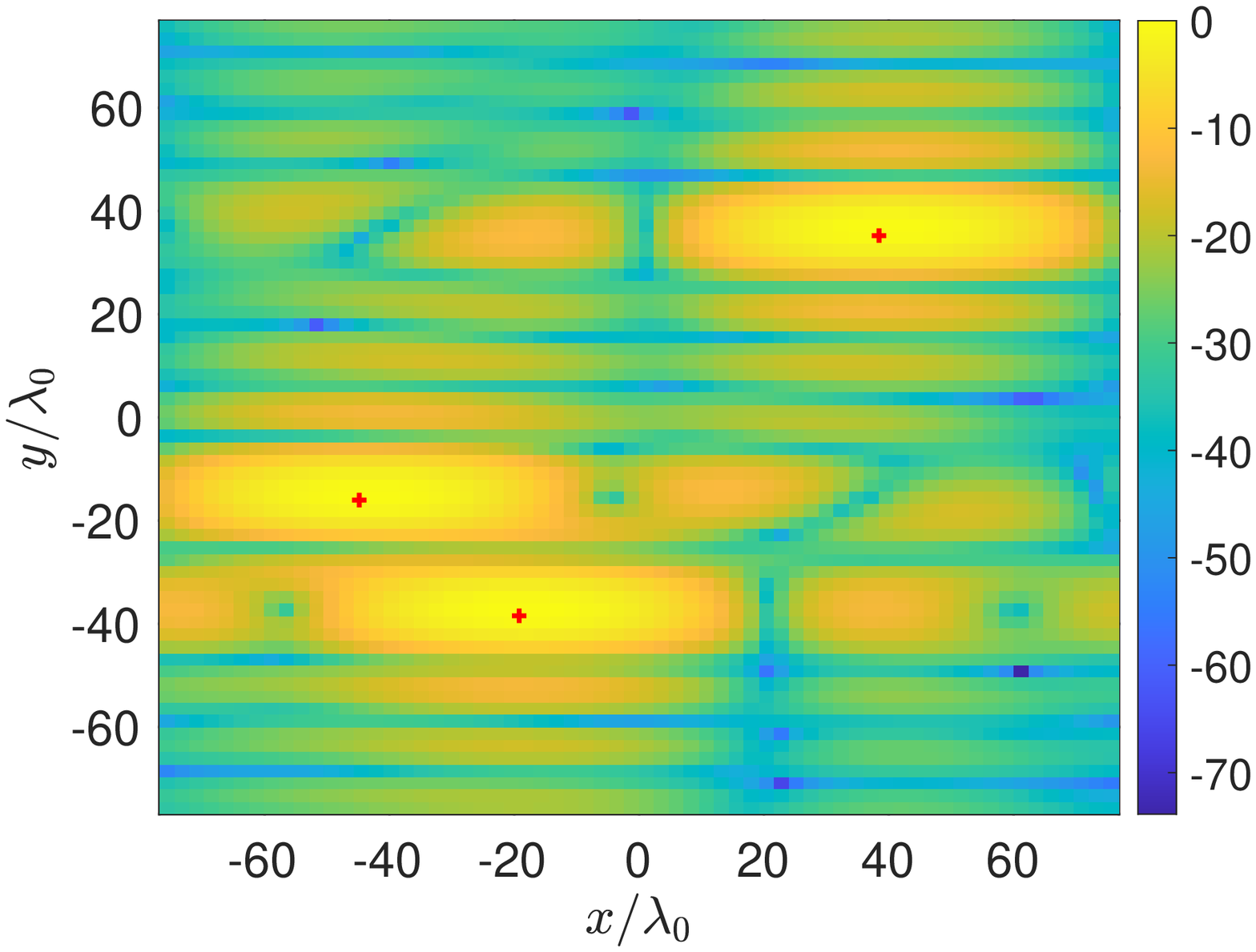}
  \hfill
  \includegraphics[width=0.48\linewidth]{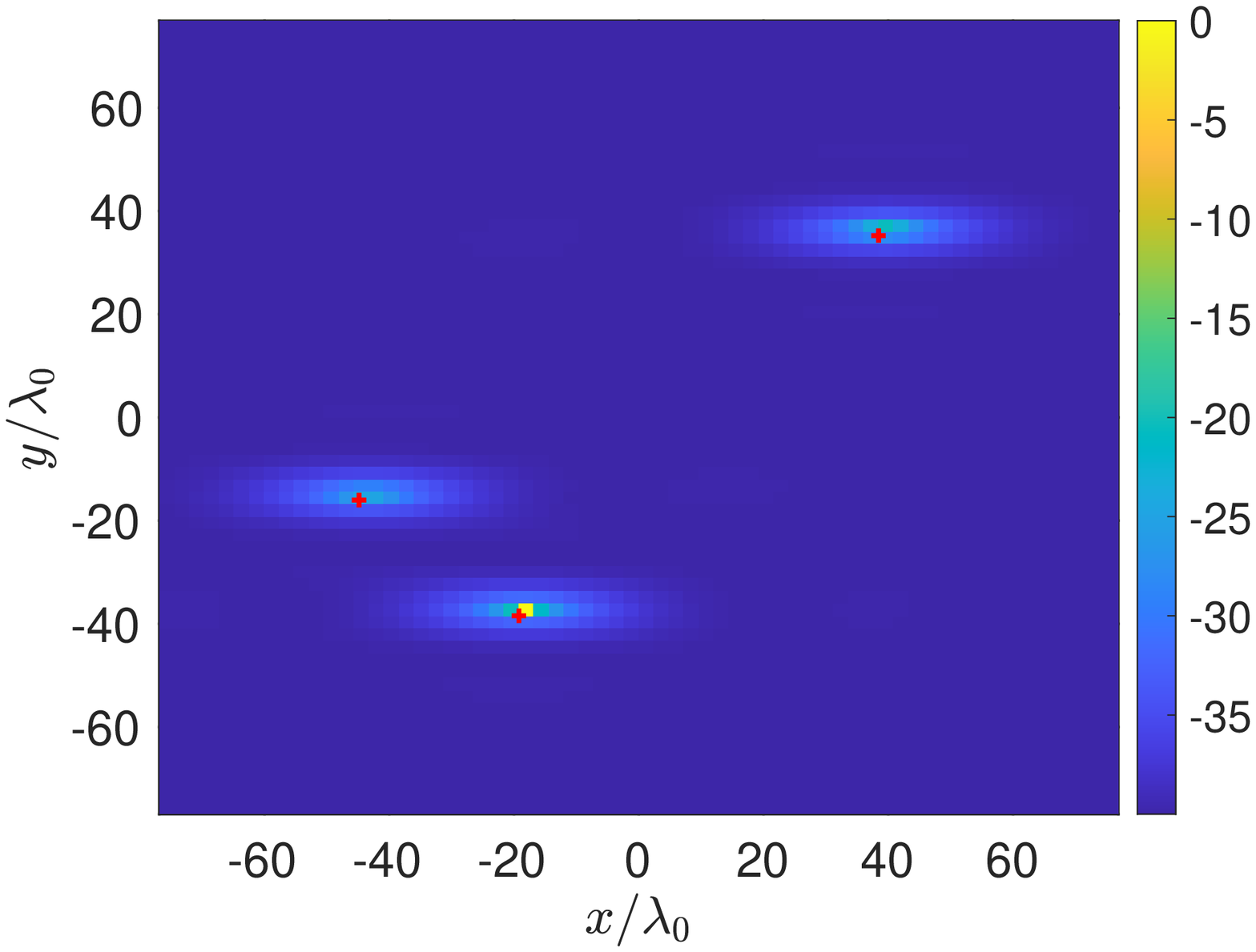}
  \caption{(Left) Image formed by KM normalized by its peak value and
    plotted in dB ($10 \log_{10}$-scale) with
    $\text{SNR} = 14.1217\, \text{dB}$. (Right) Image formed using
    modified KM with $\epsilon = 10^{-4}$ for the same measurement
    data.}
  \label{fig:3targets}
\end{figure}

We now consider an imaging scene with three identical targets. The
targets are located at
$\ypos_{1} = ( -1.4\, \text{m}, -0.5\, \text{m}, 0\, \text{m})$,
$\ypos_{2} = ( -0.6\, \text{m}, -1.2\, \text{m}, 0\, \text{m})$, and
$\ypos_{3} = ( 1.2\, \text{m}, 1.1\, \text{m}, 0\, \text{m})$. The
image formed using KM is shown in the left plot and the imaged formed
using modified KM is shown in the right plot of
Fig.~\ref{fig:3targets}. Measurement noise was added to the signals so
that $\text{SNR} = 14.1217\, \text{dB}$. The KM image shows peaks at
the target locations indicated as red ``$+$'' symbols. However, the
interactions of the side-lobes yields several imaging artifacts. In
contrast, the modified KM image is free from imaging artifacts and
identifies sharp regions where the targets are located.

A challenge in using the modified KM is that it relies on an exact
cancellation at a peak value that is normalized to one. When there are
multiple targets and the KM image is normalized by its maximum value,
presumably on only one of those targets incurs this exact
cancellation. Consequently, the image of the other targets may not be as
sharp.  However, using the image shown on the right plot of
Fig.~\ref{fig:3targets}, one can identify small sub-regions where each
of the three targets are located. If we were to evaluate the modified KM
in each of these sub-regions, we expect to obtain high-resolution
images of the individual targets because each of those images would be
normalized by the maximum value in the small sub-region corresponding
to the individual target. 

\begin{figure}[htb]
  \centering
  \includegraphics[width=0.32\linewidth]{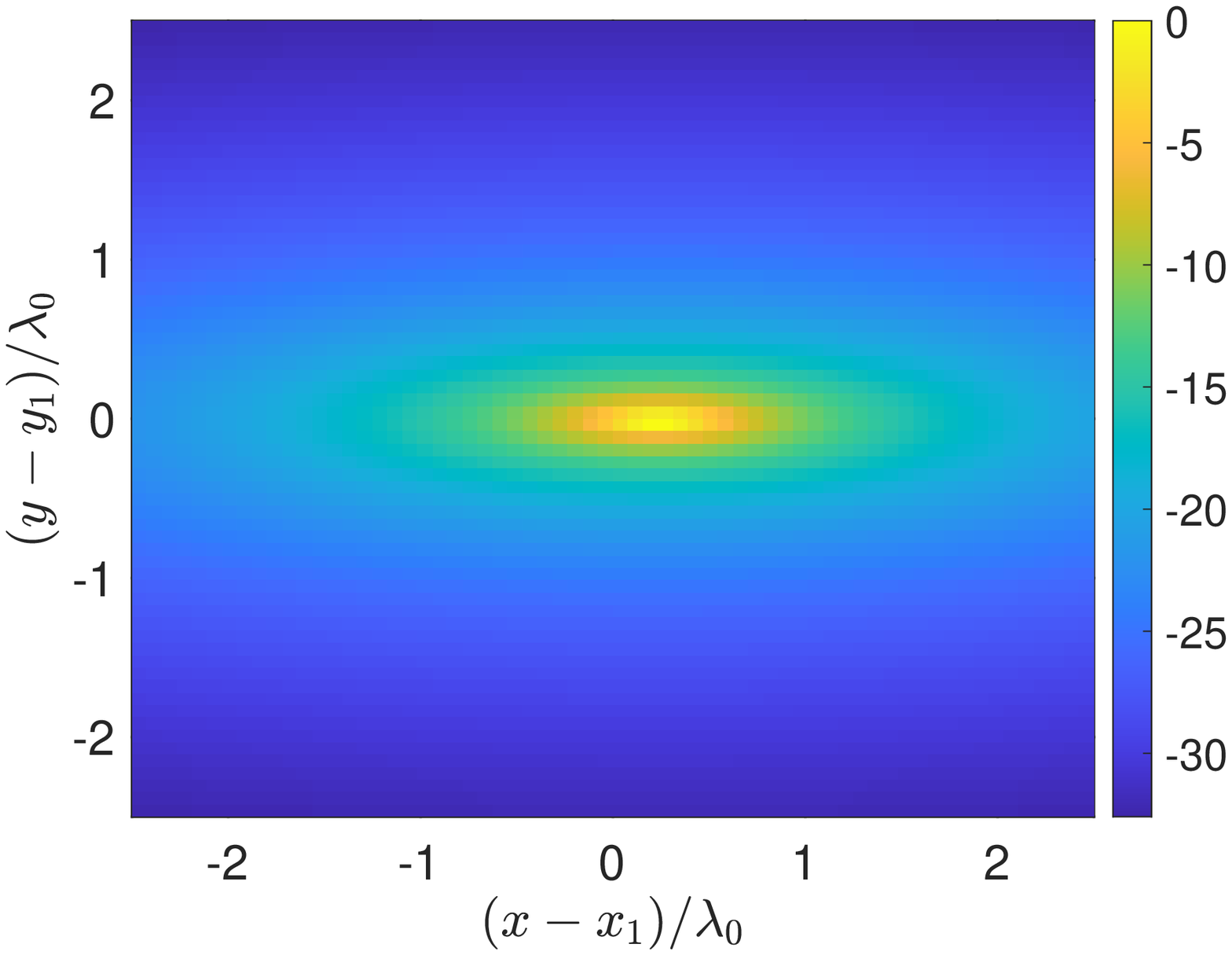}
  \hfill
  \includegraphics[width=0.32\linewidth]{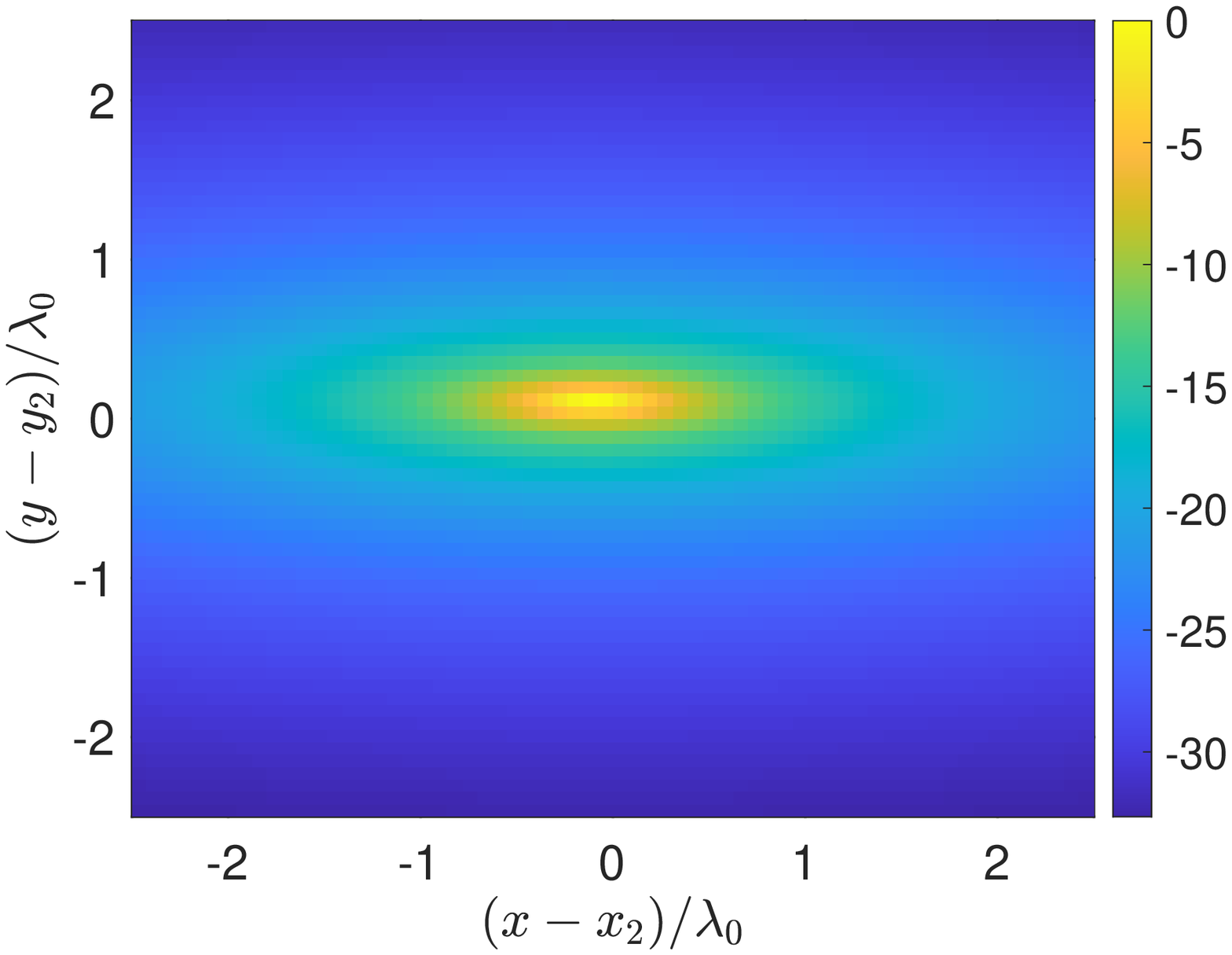}
  \hfill
  \includegraphics[width=0.32\linewidth]{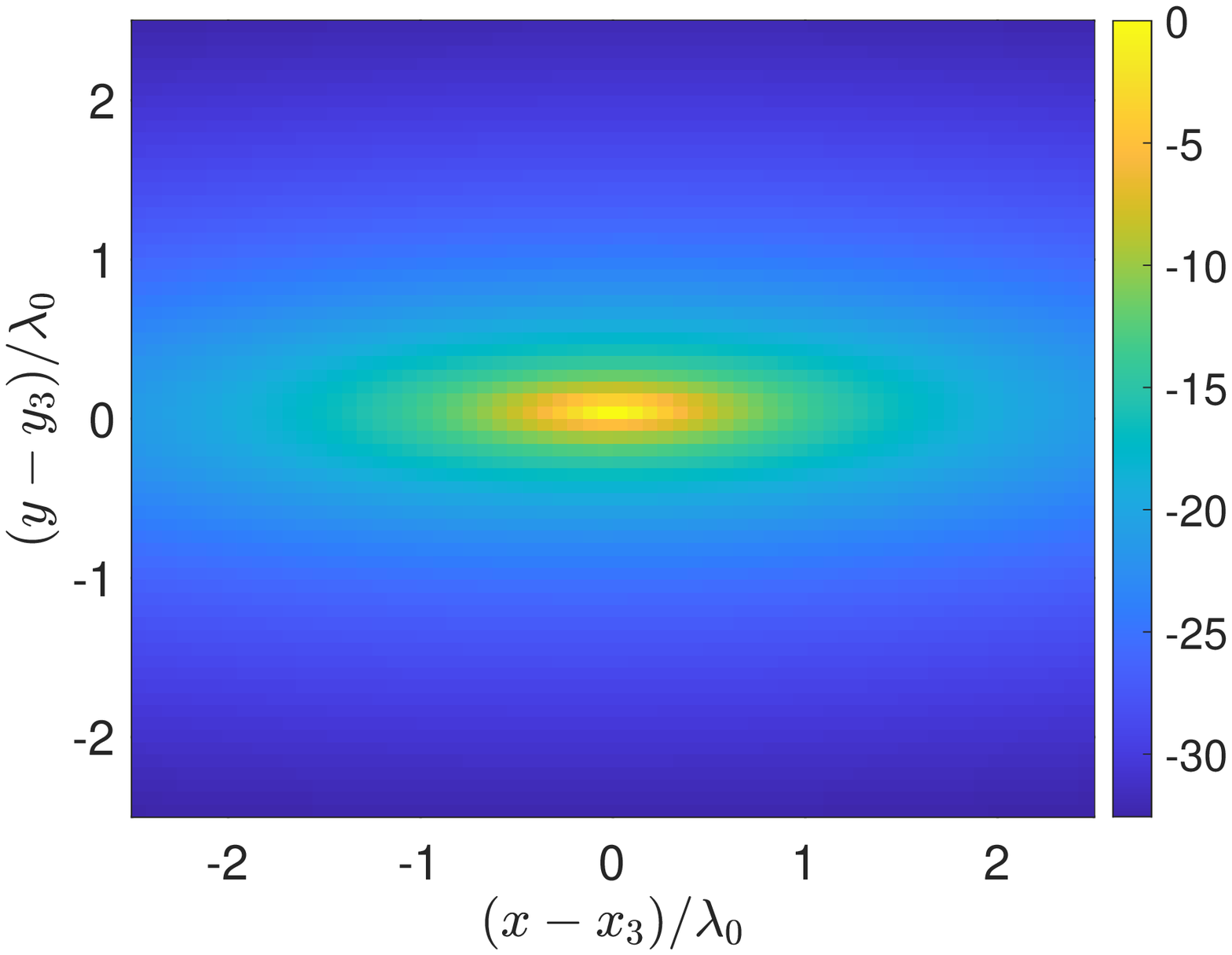}
  \caption{Close-up images of target 1 (left), target 2 (center), and
    target 3(center) in a small $5 \lambda_{0} \times 5 \lambda_{0}$
    sub-region about the target location using the same data used in
    Fig.~\ref{fig:3targets}. Each image is centered about the true
    target location and normalized by its maximum value.}
  \label{fig:3targets-closeups}
\end{figure}

In Fig.~\ref{fig:3targets-closeups} we plot close-up images in a
$5 \lambda_{0} \times 5 \lambda_{0}$ sub-region centered about the
true location of target 1 (left), target 2 (center), and target 3
(right) using $\epsilon = 10^{-4}$. The images are formed using the
same data used in Fig.~\ref{fig:3targets}. The only difference
is evaluating the modified KM only in a small sub-region.  These images
demonstrate the very high resolution achievable using modified
KM. Moreover, the resolution achieved by this method is set by the
user-defined parameter $\epsilon$.

\begin{figure}[htb]
  \centering
  \centering
  \includegraphics[width=0.48\linewidth]{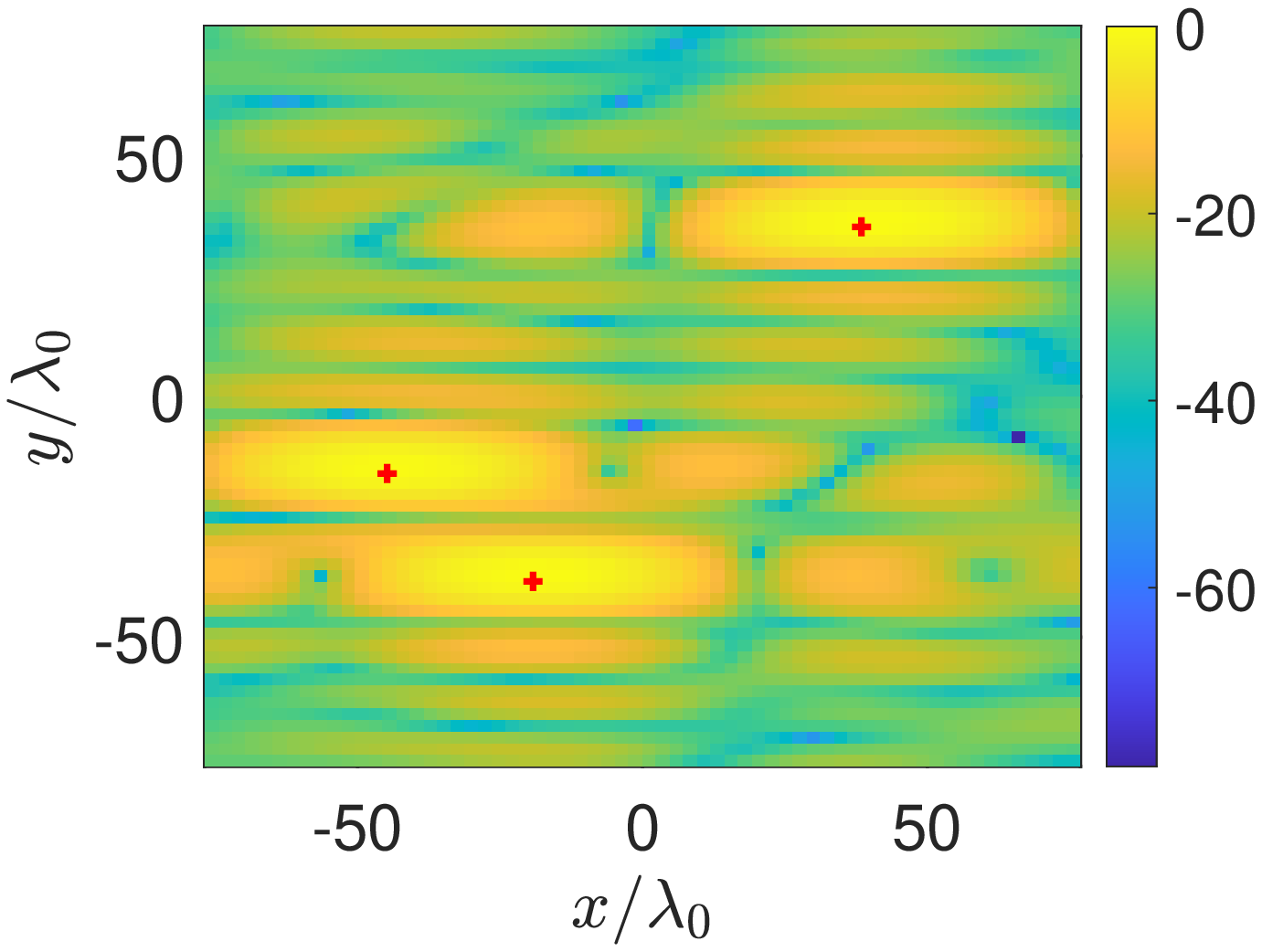}
  \hfill
  \includegraphics[width=0.48\linewidth]{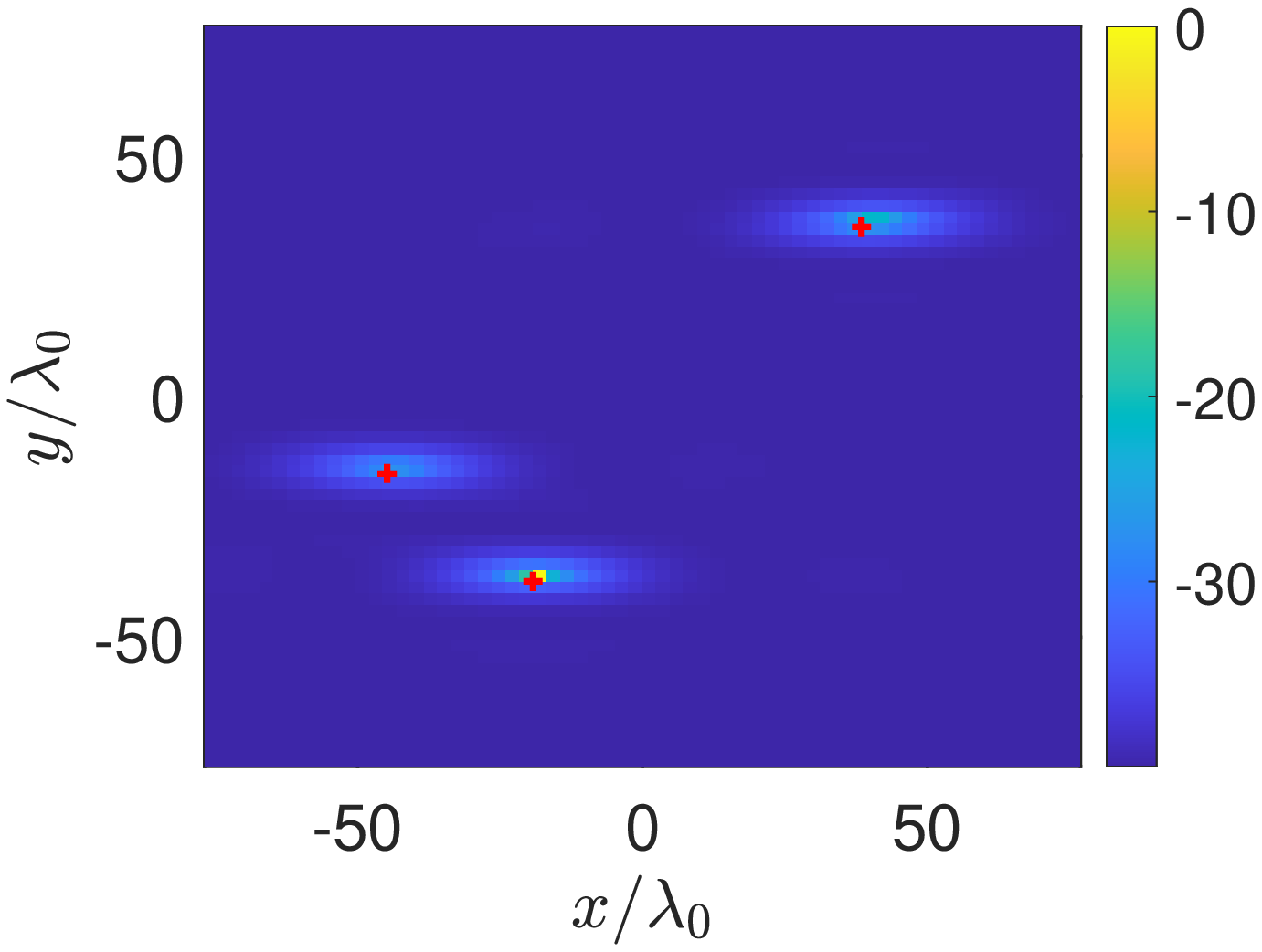}
  \caption{(Left) Image formed by KM normalized by its peak value and
    plotted in dB ($10 \log_{10}$-scale) with
    $\text{SNR} = 4.1217\, \text{dB}$. (Right) Image formed using
    modified KM with $\epsilon = 10^{-4}$ for the same measurement
    data.}
  \label{fig:3targets-lowSNR}
  \includegraphics[width=0.32\linewidth]{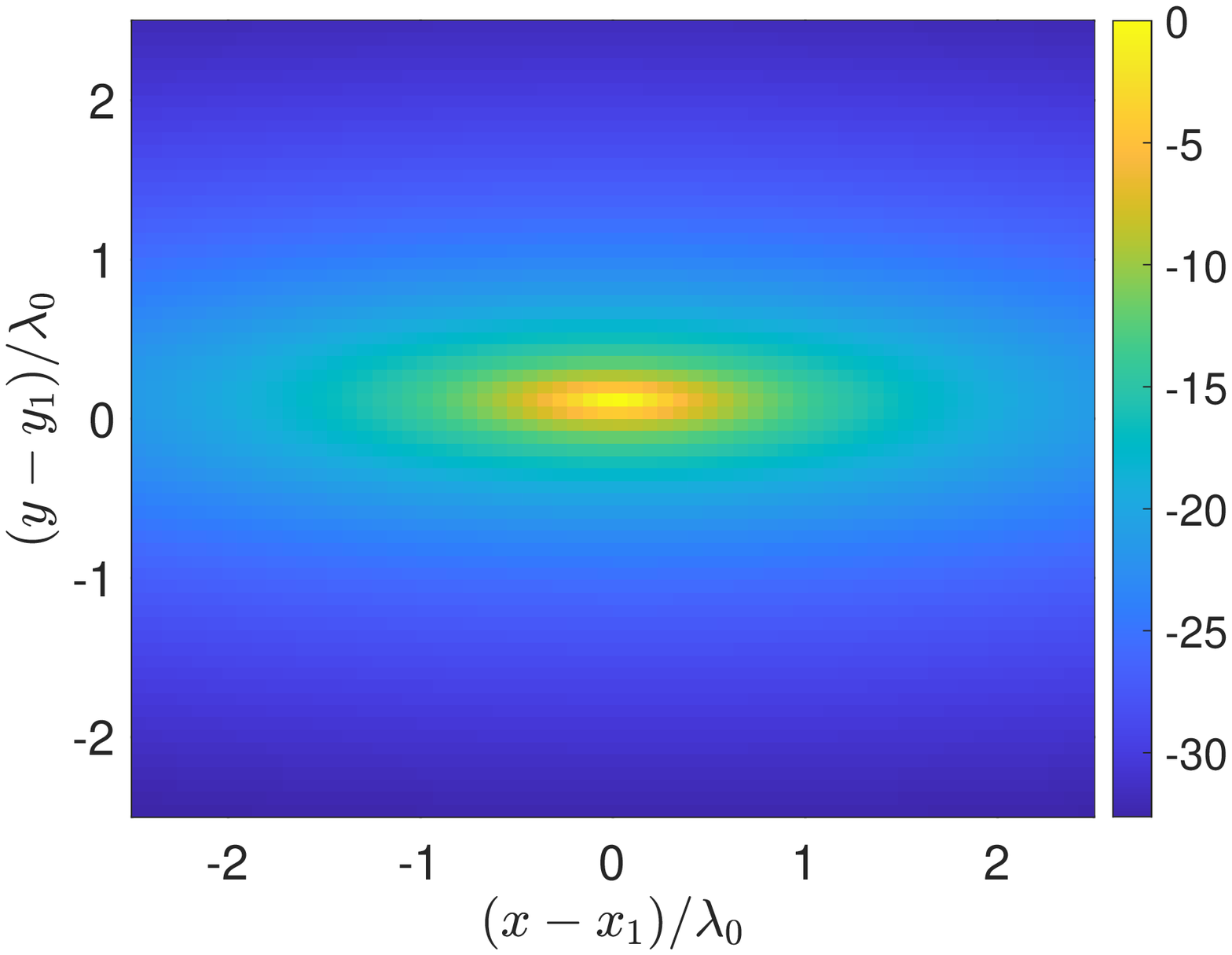}
  \hfill
  \includegraphics[width=0.32\linewidth]{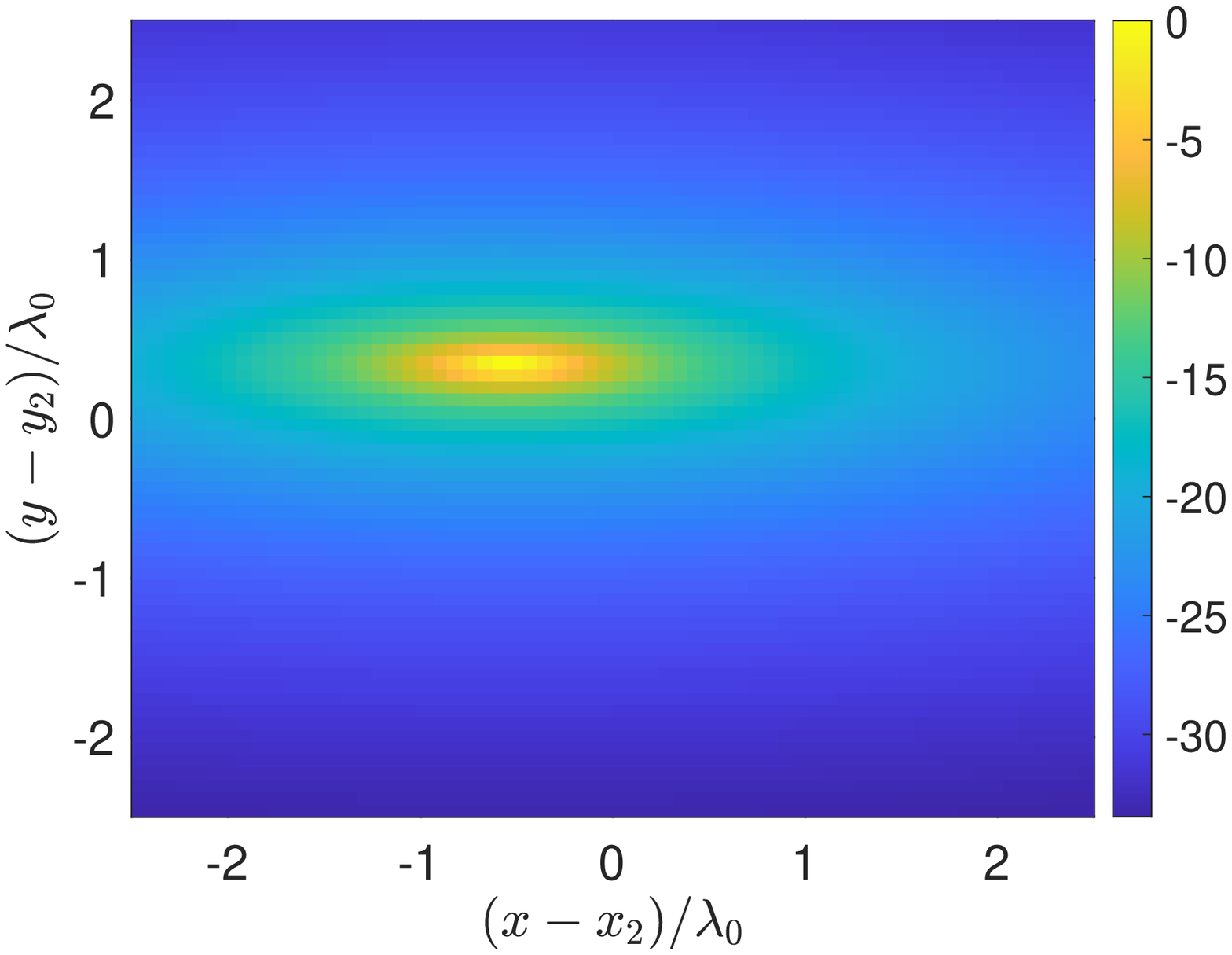}
  \hfill
  \includegraphics[width=0.32\linewidth]{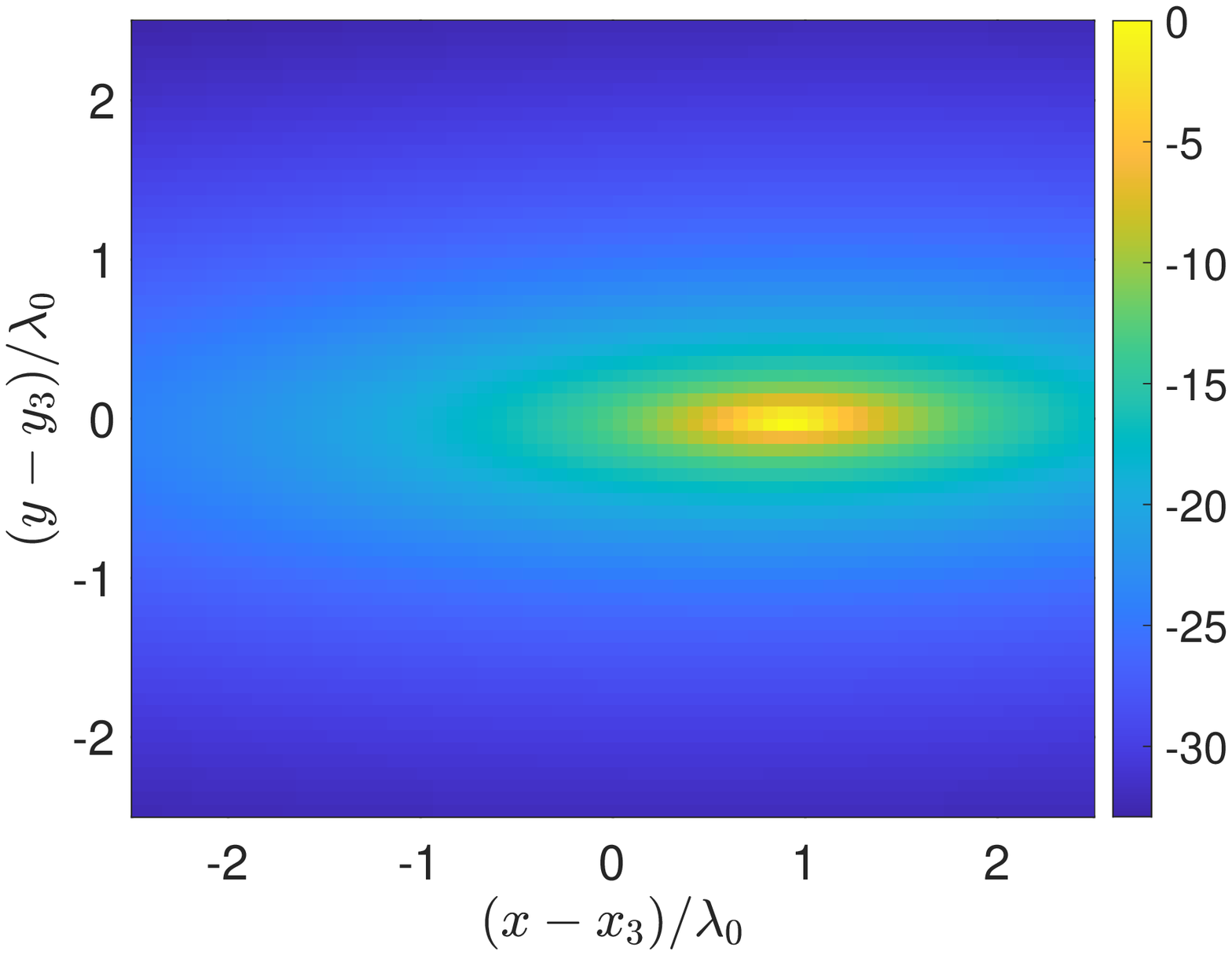}
  \caption{Close-up images of target 1 (left), target 2 (center), and
    target 3(center) in a small $5 \lambda_{0} \times 5 \lambda_{0}$
    sub-region about the target location with
    $\text{SNR} = 4.1217\, \text{dB}$.}
  \label{fig:3targets-closeupslowSNR}
\end{figure}

The results in Fig.~\ref{fig:3targets-closeups} show that the images
are not peaked at the exact target locations. They are misaligned by
fractions of a wavelength. Because modified KM yields such
high-resolution images of targets compared with KM, we find that it
shows how noise affects KM, and hence modified KM. The images produced
by KM and modified KM when we increase the noise such that
$\text{SNR} = 4.1217\, \text{dB}$ are shown in
Fig.~\ref{fig:3targets-lowSNR}.  The relative performance of both of
these imaging methods appears to be the same as for the higher SNR
case shown in Fig.~\ref{fig:3targets}. However, the plots of close-up
images about each of the target locations shown in
Fig.~\ref{fig:3targets-closeupslowSNR} using the high-resolution
modified KM images show that the peaks are farther away from the true
target locations than for the higher SNR case shown in
Fig.~\ref{fig:3targets-closeups}, on the order of a wavelength. We see
that this effect is more pronounced in cross-range than in range.

With very low SNR, the noise affects the phase of the measured signals
in the form of random phase rotations. These phase rotations shift the
predicted location of targets slightly. When plotting images using KM,
the image resolution is too coarse to be able to observe these shifts
in target locations. However, with modified KM, the resolution is so
high that one observes these shifts. Our simulation results suggest
that this random shifts in the target locations are bounded by a
wavelength or less.

\section{Conclusions}
\label{sec:conclusions}

We have identified the fundamental mechanism leading to tunable
high-resolution images with our generalization of MUSIC. Using that
same mechanism we have introduced a simple modification to KM. This
modification is a rational transformation of the
normalized KM image that includes a user-defined parameter, which we
call $\epsilon$, that effectively tunes the resolution.  We have shown
that the resolution of the modified KM method is
$O(\sqrt{\epsilon} c/B)$ in range and
$O(\sqrt{\epsilon} \lambda_{0} L/a)$ in cross-range. Because this
user-defined parameter can be made arbitrarily small, we can achieve
sub-wavelength resolution of targets using this method. In contrast to
MUSIC, this modified KM method can be applied to measurements with low
SNR.

When SNR is very low, the modified KM image of targets is shifted from
their true location. We are able to observe these shifts because the
modified KM produces such high resolution images. These shifts in the
target locations are due to noise causing phase perturbations to
measurements. In our simulations, we have observed that when
$\text{SNR} \approx 0\, \text{dB}$, these shifts are on the order of a
wavelength.

Because this modified KM method can be applied with no additional cost
beyond the KM method, itself, and because it produces tunable,
high-resolution images, we believe that this method is very useful for
a broad variety of SAR imaging problems.

\section*{Acknowledgments}

The authors acknowledge support by the Air Force Office of Scientific
Research (FA9550-21-1-0196). A.~D.~Kim also acknowledges support by
the National Science Foundation (DMS-1840265).

\bibliographystyle{abbrv}
\bibliography{MUSIC+SAR}

\end{document}